\documentclass[a4paper,aps,pra,twocolumn,superscriptaddress]{revtex4}
\usepackage{amssymb}
\usepackage{amsmath}
\usepackage{amssymb}
\usepackage{amsfonts}
\usepackage{amsthm}
\usepackage{latexsym}
\usepackage{hyperref}
\usepackage{cleveref}
\usepackage{graphicx}

\hypersetup{
    colorlinks,
    citecolor=blue,
    linkcolor=blue,     urlcolor=blue
}

\newcommand{\diffvec}[1]{\mathrm{d}\mathbf{#1}}

\begin{document}

\title{Amplified short-wavelength light scattered by relativistic electrons in the laser-induced optical lattice}
\author{I.A. Andriyash}
\affiliation{Laboratoire d'Optique Appliqu\'ee, ENSTA-ParisTech, CNRS, Ecole Polytechnique, UMR 7639, 91761 Palaiseau,
France}
\affiliation{P. N. Lebedev Physics Institute, Russian Academy of Sciences, Moscow 119991, Russia}
\email{igor.andriyash@gmail.com}
\author{V.T. Tikhonchuk}
\affiliation{Univ. Bordeaux, CNRS,  CEA, CELIA, UMR 5107,  F33400 Talence,
France}
\author{V. Malka}
\affiliation{Laboratoire d'Optique Appliqu\'ee, ENSTA-ParisTech, CNRS, Ecole Polytechnique, UMR 7639, 91761 Palaiseau,
France}
\author{E. D'Humi\`eres}
\author{Ph. Balcou}
\affiliation{Univ. Bordeaux, CNRS,  CEA, CELIA, UMR 5107,  F33400 Talence,
France}


\begin{abstract}

The scheme of the XUV / X-ray free electron laser based on the optical undulator created by two overlapped transverse
laser beams is analyzed. A kinetic theoretical description and an \textit{ad hoc} numerical model are developed to
account for the finite energy spread, angular divergence and the spectral properties of the electron beam in the optical
lattice. The theoretical findings are compared to the results of the one- and three-dimensional numerical modeling with
the spectral free electron laser code PLARES.

\end{abstract}

\maketitle

\section{Introduction}\label{sec0}

The powerful sources of X-rays are now becoming the indispensable tools in science, technology and medicine. Nowadays,
the high quality X-ray and XUV light is delivered by the synchrotron radiation (SR) sources. These installations are
typically large-scale and include a radio-frequency electron accelerator combined with an undulator
assembled from the permanent magnets, and the devices for the beam manipulation (quadruples, chicanes, steerers etc).
The last generation of SR sources -- the X-ray free electron lasers (XFEL) -- have reached the record GW powers for the 
\aa{}ngstr\"{o}m wavelengths operating in the coherent regime \cite{emma:Nat2010}. This regime is provided by the the
high stability of beam-undulator interaction, when the individually emitting electrons get progressively involved into
the process of stimulated scattering and amplify the light in the collective fashion \cite{mcneil:NatPhot2010}. 

The wavelength of SR produced by an individual electron is $\lambda_s \simeq
(1+K_0^2/2)\lambda_u/2\gamma_e^2$, where $\lambda_u$ and $K_0$ are the undulator period and its strength parameter, and
$\gamma_e$ is the electrons Lorentz factor. The parameter $K_0$ defines the efficiency of the undulator, i.e.
its capability to deviate the particles transverse momentum. $K_0$ is typically proportional to the field amplitude and
the period, e.g. for a conventional linear magnetic undulator $K_0 = 0.93 B_0 \text{[T]} \lambda_u
\text{[cm]}$. Therefore the maximal X-ray photon energy $2\pi c\hbar/\lambda_s$, is limited by the minimal $\lambda_u$
and the energy of the electron. In the conventional SR sources, the size of the undulator magnet which provides
sufficiently strong magnetic field is around centimeter, which makes it difficult to reach the sub-\aa{}ngstr\"{o}m
radiation wavelengths, requiring larger and more expensive accelerators.

Alternatively, a number of  currently explored schemes propose to collide electron beam with  strong optical laser
radiation \cite{albert:PRSTab2010,taphuoc:NatPhot12}. Such Compton sources can  provide a very short, micrometer
undulation period with  sufficiently large $K_0$ values. The rate of the amplification in FELs is typically
determined by the parameter $\rho \propto \gamma_e^{-1} j_e^{1/3} (K_0 \lambda_u)^{2/3}$, where $j_e = I_e/\sigma_x^2$
is the peak flux defined by the current $I_e$ and the transverse size of the electron beam $\sigma_\perp$
\cite{huang:PRSTAB2007}. Since amplification decreases with $\lambda_u$, the efficient operation of the optical-based
XFELs requires a relatively long interaction distances which imposes severe limitations on the beam quality.  As a
consequence, while such an idea is being discussed in a number of theoretical works
\cite{sprangle:JAP1979,Bacci:PRSTAB2006,sprangle:PRSTAB2009}, any practical realization remains beyond current
expectations.

One possible solution to improve the concept of the optical-based XFEL, is to replace the conventional RF linac with the
laser plasma accelerator (LPA) \cite{esarey:RMP2009,malka:POP2012}. The technology of LPA has already proved its
capability to deliver  pico-Coulomb femtosecond beams of MeV electrons \cite{faure:Nat2006}, and has now reached the
GeV level \cite{kim:PRL2013,leemans:PRL2014}. The incoherent undulator radiation produced by the laser accelerated
electrons was recently observed experimentally \cite{schlenvoigt:Nat2008,fuchs:Nat2009} and the prospects for 
coherent amplification of this radiation were discussed \cite{nakajima:NatPhys2008}. Although the laser plasma
accelerators now provide the necessary high electron flux, the level of collimation and monoenergeticity required for
the X-rays amplification yet remains challenging.

In our work we tackle this problem considering a scheme based on a specific configuration of the optical undulator --
the optical lattice. The lattice results from the overlap of two identical side laser beams and represents a spatially,
transversally modulated electromagnetic field structure. The interest for such a structure and its interaction with
electron beams originates from the works of P.L. Kapitza and P.A.M. Dirac back in 1933 \cite{kapitza:MPCPS1933}, and was
later revised in a number of works \cite{bucksbaum:PRL88,Freimund:Nat2001}. At the same time appeared
the idea to use the electron slow motion  in the optical lattice as a low frequency light source
\cite{fedorov:APL1988,sepke:PRE2005}. Recently we have focused our attention to electron light scattering in the lattice
for coherent amplification  of short wavelength radiation \cite{balcou:EPJD2010,andriyash:PRL2012}.

A major advantage of the transverse optical lattice is that it both wiggles the electrons in the varying laser field,
and its spatial modulations act on the electrons via the ponderomotive force thereby trapping them in the potential
channels \cite{andriyash:EPJD2011,frolov:NIMB2013}. On one hand, such electron guiding prevents electrons from diverging
along one direction \cite{andriyash:PRSTAB13}, thus, partially preserving the electron flux. On the other hand,
potential channels affect the collective behavior of the electrons predisposing them to a new mechanism of amplification
similar to the stimulated Raman scattering \cite{andriyash:PRL2012}. Similarly to the traveling-wave technique proposed
for the Thomson scattering in \cite{balcou:EPJD2010,debus:APB2010,steiniger:NIMA2014}, the optical lattice can
co-propagate with electrons for a long distance and therefore may  provide a stable amplification.

In the present work, we revise the physics of X-rays amplification in the electron beam trapped in the optical lattice.
For this we develop a self-consistent kinetic model in \cref{sec1}, and with its help describe the growth of
electromagnetic signal in the analytical approach and by using a simple numerical integration (see \cref{sec2}). In
\cref{sec3}, the theoretical results are compared to the three-dimensional numerical modeling performed with the
unaveraged spectral free electron laser code PLARES \cite{andriyash:JCP2014}. The concluding remarks are given in
\cref{sec4}.

\section{Theoretical model}\label{sec1}

We describe interaction of the relativistic electrons with the optical lattice and the scattered radiation. The
theoretical model is based on the Vlasov kinetic equation and the equation for electromagnetic potential:
\begin{subequations}\label{set0}
 \begin{equation}\label{set0:1}
 \partial_t f + (\mathbf{v} \nabla) f + (\partial_t\mathbf{a} - \mathbf{v}\times\nabla\times\mathbf{a})\,
\partial_\mathbf{p} f = 0\,,
 \end{equation}
 \begin{equation}\label{set0:2}
  (\partial_t^2 - \nabla^2)\mathbf{a} = - \int \mathbf{v} f \diffvec{p} \,.
 \end{equation}
\end{subequations}
In present study, we normalize frequencies and wavenumbers to $\omega_0 = 2\pi c/\lambda_0$ and $k_0 = \omega_0/c$,
where $\lambda_0$ is the wavelength of the laser. The distances and time in \cref{set0} are measured in the units
$c/\omega_0$, $\omega_0^{-1}$, the electron velocities and momenta are normalized to the speed of light $c$ and
$m_e c$, respectively, and the vector potential is in the units of $m_ec^2/e$. In this convention, the electron
distribution function (EDF) is normalized as $\int f(\mathbf{r},\mathbf{p},t) \diffvec{p} = n_e$, where the electron
density $n_e$ is in the units of its critical value, $n_c = (m_e\omega_0^2/4\pi e^2)$. 

\begin{figure}[h!]\centering
\includegraphics[width=0.3\textwidth]{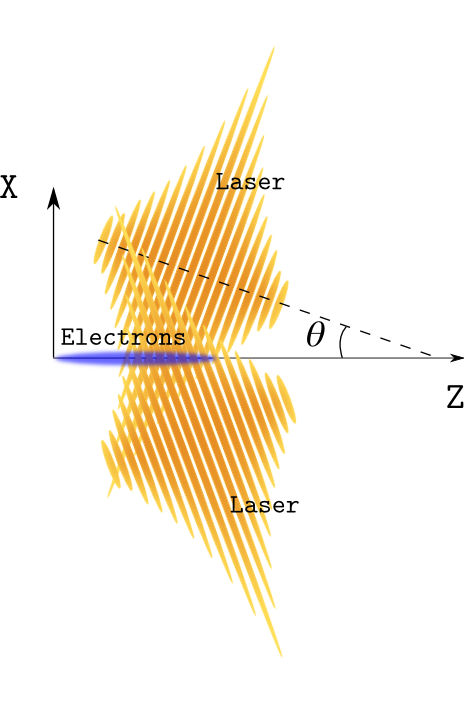}
\caption{Optical lattice scheme}\label{fig1}
\end{figure}

The beam of electrons travels at a relativistic velocity, $v_b = \sqrt{1-\gamma_b^{-2}}\simeq 1$, in the overlap of two
identical laser pulses incident symmetrically with the angle $\theta$ (see \cref{fig1}). The vector potential
$\mathbf{a} = \mathbf{a}_L + \mathbf{a}_s$ in \cref{set0} accounts for the lattice field,
\begin{subequations}\label{set1}
\begin{equation}\label{lattice:0}
 \mathbf{a}_L = 2 a_0 \mathbf{e}_y \sin(x \sin\theta) \cos(t-z\cos\theta)\,,
\end{equation}
and the signal wave $\mathbf{a}_s$, associated with the scattered radiation. This signal co-propagates with electrons:
\begin{equation}\label{signal:0}
 \mathbf{a}_s = a_s\mathbf{e}_y \cos(k_s(t-z))\,,
\end{equation}
\end{subequations}
and its amplitude $a_s$ is assumed to be small, so in our analysis we retain only linear terms on $a_s$.

One approximation commonly used to describe the motion of the charged particles in a strong laser field, is to divide
this motion into the ``fast'' and ``slow'' parts. The fast reaction of the electrons is to follow oscillations of
the electromagnetic field directly. On the other hand, the particles are also driven by the period-averaged
ponderomotive force, defined by the field spatial gradients. The optical lattice field in \cref{lattice:0} has the 
transverse gradient, which creates the ponderomotive force along $x$-axis:
\begin{subequations}\label{set2}
\begin{equation}\label{ponder:0}
 \mathbf{F}^{(0)} = -\mathbf{e}_x \gamma_b^{-1} a_0^2\partial_x|\sin(x \sin\theta)|^2  \,,
\end{equation}
and produces a series of the potential channels with the widths $L_x = \pi/2\sin\theta$.

In presence of the signal wave (\ref{signal:0}), its interference with the lattice produces a beat-wave in the
longitudinal direction. This wave acts on the electrons via the ponderomotive force:
\begin{align}\label{ponder:1}
 \mathbf{F}^{(1)} &= \mathbf{e}_z a_0 a_s\gamma_b(k_s(1-v_b)+v_b-\cos\theta)\\
 &\sin(x \sin\theta) \sin\Big[ (1-k_s)t-(\cos\theta-k_s)z\Big]  \nonumber\,.
\end{align}
\end{subequations}
which is calculated in the reference frame of electron beam (cf. \cite{gibbon:2005}), and then translated to the
laboratory system.

\subsection{Unperturbed beam in the lattice}\label{sec1:1}

The unperturbed state of the electron beam trapped in the lattice potential is defined only by the transverse
ponderomotive force \cref{ponder:0}. The characteristics of the kinetic equation (\ref{set0:1}) in this case can be
written as:
\begin{equation}\label{character:0}
 \frac{\mathrm{d}x}{v_x} = \frac{\mathrm{d}p_x}{F^{(0)}}\,,\quad \mathrm{d}t = \frac{\mathrm{d}z}{v_z}\,,
\end{equation}
and they define the particles trajectories, along which the initial EDF remains constant. The first equation of
\cref{character:0} represents the energy conservation condition,
\begin{equation}\label{conserv:1}
 p_x^2/2a_0^2 + \sin^2(x \sin\theta) = \xi^2\,,
\end{equation}
and, neglecting the variations of the electron Lorentz factor, we may calculate the orbits of the electrons as:
\begin{equation}\label{orbits:1}
 \sin(x\sin\theta) = \mathrm{sn}(\xi\Omega_0 t+\Theta; \xi^{-2})\,.
\end{equation}
where $\mathrm{sn}(u,m) \equiv \sin(\phi)$ is a Jacobi elliptic function defined as an inverse of elliptic integral
${u = \int_0^\phi (1-m\sin^2 \theta)^{-1/2}\, \mathrm{d}\theta}$. Here we have also denoted a normalized excursion
$\xi$,
and $\Omega_0 = \sqrt{2}a_0\sin\theta/\gamma_e$ is a frequency of the small-amplitude oscillations. The phase $\Theta$
is defined by the initial coordinates and momentum of the particle. 

If $\xi<1$, the electron is trapped, and it oscillates in a channel of the optical lattice with the frequency:
\begin{equation}\label{betatron_freq}
 \Omega = \cfrac{\pi\,\Omega_0}{2 K(\xi)}\simeq \Omega_0 \left(\cfrac{\sin\pi \xi}{\pi
\xi}\right)^{1/6} \approx\Omega_0 (1-\xi^2/3)\,,
\end{equation}
where $K(m) = \int_0^{\pi/2}  (1-m\sin^2 \phi)^{-1/2}\, \mathrm{d}\phi$, is an quarter period of elliptic integral. 

The first approximation in \cref{betatron_freq} was previously presented in \cite{andriyash:PRSTAB13}, and its accuracy
is better than 99\% for $\xi<0.999$. The alternative expression (in the leftmost part of \cref{betatron_freq}) is
derived by expanding $\sin^2(x \sin\theta)$ into the series, and it retains a 98\% accuracy for $\xi<0.85$. The latter
is more convenient for the analytical developments, and we will use it further with the electron 
orbits \cref{orbits:1} approximated by sinusoids. Finally, the set of electron trajectories in the $(x\,,z)$
phase-plane
will read:
\begin{align}\label{trajects}
 &X = \Big(x_0 \cos(\Omega t) + p_{x0}/(\Omega\gamma_e) \sin(\Omega t) \Big)\,,\nonumber\\
 &P_x =  \Big((p_{x0} \cos(\Omega t) - x_0\Omega\gamma_e \sin(\Omega t)\Big)\,,\\
 & Z = z_0 + v_z t\,,\quad v_z =\mathrm{const}\,. \nonumber
\end{align}

For a simple model analysis one can assume homogeneous electron distributions in the transverse phase plane, where the
particles are trapped, and in the longitudinal direction, which is not affected by the lattice field. The electron
distribution function in this case reads:
\begin{equation}\label{EDF:0}
 f = \frac{n_0}{4\sqrt{2}a_0\xi_0 \delta \gamma_b} \,\eta\left(\xi_0^2 -\xi^2\right)\eta(\delta \gamma -
|p_z-p_b|)\,,
\end{equation}
where $\eta(x)$ is a unit-step function, $n_0 = n_e(x=0)$ is a maximal electron density, and $\delta \gamma$ is a
longitudinal energy spread of electrons. The maximal excursion is limited by the potential scale, $\xi_0\le 1$. 

\subsection{Interaction of the electron density perturbations and the signal electromagnetic wave}\label{sec1:2}

The force (\ref{ponder:1}) associated with the signal wave acts on the particles, and rearranges their longitudinal
positions, thus generating the perturbation of EDF $f \to f+ \delta f$. In its turn, this perturbation acts back on the
signal field through the electron current along $y$-axis in the right hand side of \cref{set0:2}. Such coupling may
provide a resonant interaction between electron and electromagnetic modes resulting in their amplification or damping
similarly to the stimulated Raman scattering \cite{Kruer:1988}.

During the early, linear stage of interaction, the amplitude of the signal wave and the EDF perturbations are small
$a_s\ll a_0$, $\delta f\ll f$. The current $\int \mathbf{v} f \diffvec{p}$, associated with the resonant interaction in
\cref{set0:2}, includes the generating term $a_L \delta n/\gamma_e$, and the term $a_s n_e /\gamma_e$, which is
responsible for the dispersion of the signal resulting from its interaction with the space-charge plasma waves. In the
simplest case, we may neglect this dispersion as well as the diffraction of the amplified wave, therefore, assuming a
plane transverse profile. Keeping only the first order resonant perturbation terms, we write the signal wave
equation as:
\begin{equation}\label{maxwell:0}
 \partial_t a_s = i\,\frac{a_0 n^*}{\gamma_b k_s}\sin(x \sin\theta)  \,,
\end{equation}
where $n^*$ is related to the electron density perturbations as
$$\delta n = \int \delta f \diffvec{p} = \mathrm{Re}\left[n^* \mathrm{e}^{i (\cos\theta-k_s) z- i
(1-k_s)t} \right]\,.$$

To study the interaction of the electromagnetic field with relativistic electrons, it is convenient to introduce the
coordinates which follow the center of the electron beam $z \to z + v_b t$. The Doppler-shifted frequencies of the
lattice and the signal wave are $\tilde{\omega}_e = (1-v_b \cos\theta)$ and $\tilde{\omega}_s = (1-v_b)k_s$
respectively, and the interaction is resonant when $\tilde{\omega}_e\approx \tilde{\omega}_s$ (we will discuss the
resonance condition in detail in \cref{sec2:1}). The ponderomotive force \cref{ponder:1} and the density
perturbations in the moving coordinate system read: 
\begin{align}\label{ponder:11} 
&F^{(1)} = \frac{a_0 a_s k_s}{\gamma_b}\, \sin(x \sin\theta)\sin\Big[\Omega_s t + k_s z\Big]\,,\nonumber\\
&\delta n = \mathrm{Re}\left[n^* \mathrm{e}^{ - i(\Omega_s t+ k_s z) }\right]\,,
\end{align}
where we assumed $k_s\gg\cos\theta$, and $\Omega_s = \tilde{\omega}_e - \tilde{\omega}_s$ is a frequency of the
slow oscillations of longitudinal ponderomotive force. Note that, it is the force $F^{(1)}$, which produces the
modulations of electron density -- the bunched structure. These modulations propagate with the velocity
$-\Omega_s/k_s$, and the sign of $\Omega_s$ defines the propagation direction. In case of $\Omega_s<0$, the modulations
co-propagate with the electron beam and accelerate the particles, thereby draining the energy of electromagnetic field.
Otherwise, if the signal is down-shifted, $\tilde{\omega}_s<\tilde{\omega}_e$, the electrons give the energy to the
wave amplifying it. In what follows, we will consider only this growing mode, always assuming $\Omega_s>0$.

Let us describe the dynamics of $n^*$ by perturbing the kinetic equation (\ref{set0:2}) and averaging it over the
lattice field period:
\begin{equation}\label{vlasov:1}
 \left[\partial_t + v_x \partial_x + v_z \partial_z +  F^{(0)} \partial_{p_x}\right] \delta f = -F^{(1)}
 \partial_{p_z}f\,,
\end{equation}
where only the first order terms are retained. The operator in the left hand side is simply the time derivative
$\mathrm{d}\delta f/\mathrm{d}t$ calculated along the trajectories (\ref{trajects}). Substituting these trajectories
into \cref{vlasov:1} and considering the model EDF \cref{EDF:0}, the integration of \cref{vlasov:1} over the momenta
may be simplified as:
\begin{align}\label{density:1} 
 n^* =  \frac{a_0 n_0 k_s^2}{4\gamma^4 \xi_0} \sin(x \sin\theta)\int_{-\infty}^{t} \mathrm{d}t' a_s(t') G(t'-t)
\end{align}
where we denote $\chi = x\sin\theta$, and the function:
$$ G(t)  = \frac{\sin(k_s\delta v_z t)}{k_s\delta v_z} \,\int_0^{\xi_0^2} \;\mathrm{d}\xi^2
\frac{\mathrm{e}^{i(\Omega-\Omega_s) t}}{\sqrt{\xi^2- \sin^2\chi}}\,,$$
describes dynamics of the electron density perturbations driven by the ponderomotive forces. Here we have also
neglected the non-resonant up-shifted mode $\exp[-i(\Omega+\Omega_s)t]$. The longitudinal
energy spread of electrons is represented here by the parameter $\delta v_z = \delta\gamma_e/\gamma_b^3$.

\section{Analysis of the stimulated scattering}\label{sec2}

\Cref{density:1,maxwell:0} describe the evolution of the coupled electron and electromagnetic perturbations. The
conditions, at which the signal wave may be amplified, are defined by the dispersion properties of the electron beam
mode. For simplicity, in the following analysis we will assume that electrons fill completely the lattice channel, so
that $\xi_0=1$.

In our model, the electron mode is described by the function $G(t)$ in \cref{density:1}. If we neglect the dependence of
$\Omega$ on $\xi$, the electron mode would be harmonic and the integral in the function $G$ may be simplified as 
\begin{equation}\label{harm_mode}
B = \int_0^1\frac{\mathrm{e}^{i \Omega t} \;\mathrm{d}\xi^2 }{\sqrt{\xi^2- \sin^2\chi}} =  \mathrm{e}^{i \Omega_0 t}
\cos\chi\,.
\end{equation}

The dependence of $\Omega$ on $\xi$ results in a deformation of the electron mode with respect to
the $x$-coordinate. Considering the second approximation in \cref{betatron_freq}, one may qualitatively expect that
the electron mode spectrum after averaging will depend on $x$ as $B \propto \mathrm{e}^{i \Omega' t}$, where $\Omega' =
\Omega_0\left(1- \sin^2\chi/3\right)$. More accurately, the mode can be calculated as 
\begin{equation}\label{ellipt_mode}
B = \sqrt{6\pi}\; t^{-1/2}\;(C(\alpha) -i S(\alpha))\; \mathrm{e}^{i \Omega' t} \,,
\end{equation}
where $S(\alpha)$ and $C(\alpha )$ are the Fresnel integrals, and ${\alpha = \cos\chi \sqrt{2t/3\pi}}$. 

\begin{figure}[ht!]\centering
\includegraphics[width=0.48\textwidth]{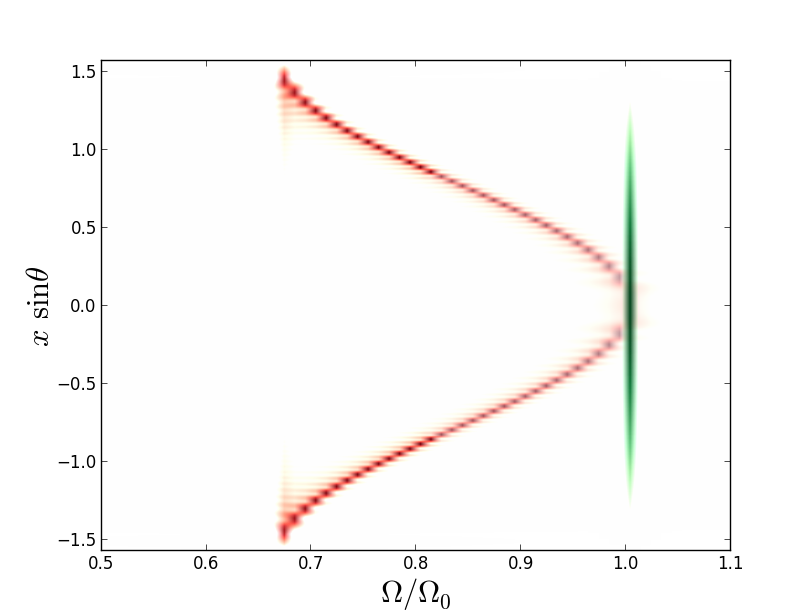}
\caption{Electron beam mode frequency as a function of $x$-coordinate for the elliptic (red) and the harmonic (green)
electron orbits.}\label{fig2}
\end{figure}

The numerically calculated spectral-spatial map of electron beam modes given
by \cref{harm_mode,ellipt_mode} is shown in \cref{fig2}. The deformation of the electron mode \cref{ellipt_mode}
shown in red may increase the spectral width of the region, where the signal wave interacts resonantly with the
electrons.

\subsection{Interaction parameters}\label{sec2:0}

For the following analysis we consider the parameters which are typical for the laser accelerated electrons. The optical
lattice field \cref{lattice:0} is produced by two plane waves corresponding to the pulses of a Ti:Sapphire laser
with the wavelength $\lambda_0 = 800$ $\mu$m and the intensity $2.16\cdot 10^{16}$~W/cm$^2$. The incidence angles of
the laser waves to the electron propagation direction are chosen to be $\theta = 10^\circ$. A 100 pC beam of 40
MeV electrons has  flat longitudinal and cylindrical transverse profiles with the duration $3.75\lambda_0/c$ and a
radius $R_b=1.3\lambda_0$, respectively. This corresponds to the dimensionless electron density and laser amplitude $n_0
= 0.0352$ and $a_0 = 0.1$. 

We also need to define the characteristics of the electron phase-space distribution. The spread of electron energies
in our tests vary from the case of monoenergetic beam $\delta\gamma=0$, to a more realistic case
$\delta\gamma/\gamma_b = 0.03$. For the numerical tests, we consider a homogeneous initial spread of the
electrons transverse momenta chosen to fit the trapping condition $\delta p_x = \sqrt{2}a_0$. In the physical units
this corresponds to 1 mrad of rms angular divergence, and the divergence along the non-trapped $y$-direction can be
assumed to be lower --  0.36 mrad.

Note that in the present analysis we do not consider the effects of the electrostatic fields. Partially this is
justified by the fact that the Coulomb repelling for the relativistic beams becomes weak with higher $\gamma_b$.
However, for a sufficiently long interaction time these effects may become important, and will be considered in 
future studies.

\subsection{Harmonic electron mode}\label{sec2:1}

Let us first study an interaction with the harmonic electron mode \cref{harm_mode}, assuming all electrons
oscillating with the same frequency $\Omega_0$. In this case, the function $G$ may be calculated analytically, and
\cref{density:1,maxwell:0} may be linearized with help of the Laplace transform, or by reducing \cref{density:1} to the
differential form and performing the time-domain Fourier transform:
\begin{align}\label{density:2} 
 &\left((\omega-\Omega_s+\Omega_0)^2-k_s^2\delta v_z^2\right) \langle n^*\rangle_\omega = \\
 &\qquad \qquad =  \langle a_s\rangle_\omega \frac{a_0 n_0 k_s^2}{2\gamma_b^4} \sin\chi \cos\chi \nonumber\,.
\end{align}
With help of the Fourier transform we can find $\langle a_s\rangle_\omega$ from \cref{maxwell:0}, and substituting it
to \cref{density:2} we obtain the dispersion equation:
\begin{equation}\label{dispersion:1} 
\left((\omega-\Omega_s+\Omega_0)^2-k_s^2\delta v_z^2\right)= \frac{a_0^2n_0\tilde{\omega}_e}{\omega\gamma^3} \sin^2\chi
\cos\chi\,. 
\end{equation}

\begin{figure}[ht!]\centering
\includegraphics[width=0.48\textwidth]{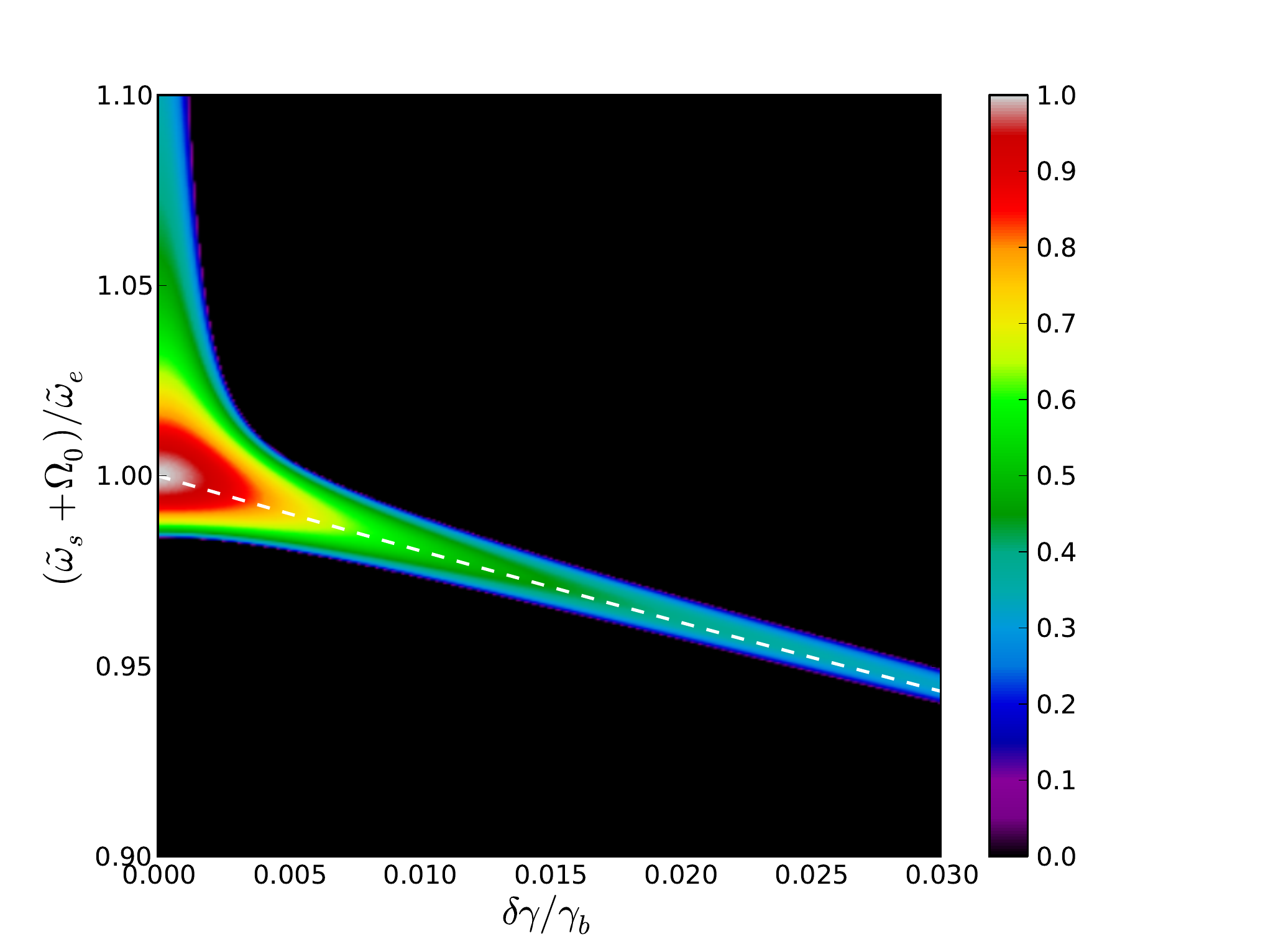}
\caption{Normalized FEL parameter $\rho/\rho_0$ as a function of longitudinal temperature and signal frequency given by
\cref{dispersion:1}. The white dashed line corresponds to the resonance condition \cref{resonance:0}.}\label{fig3}
\end{figure}

The dependence on the transverse coordinate in the right hand side of \cref{dispersion:1} corresponds to the modulation
of the coupling between the electron and electromagnetic modes. Physically this means that the signal wave interacts
with electrons more efficiently around the coupling maxima  $\chi = \pm 1/\sqrt{3}$, which tends to produce an
amplified wave with a profile modulated along the $x$-axis. For the one-dimensional analysis we consider an
effective mode coupling by averaging ${\langle \sin^2\chi \cos\chi \rangle = 2/3\pi}$. 


The coupling of the electromagnetic and electron modes in \cref{dispersion:1} allows a resonant energy
transfer from the particles to the electromagnetic field. If the solution $\omega$ has a negative imaginary part, it
corresponds to the modes which grow exponentially. It is easy to see, that for a monoenergetic beam with $\delta
\gamma =0$, the maximal growth occurs for $\Omega_s=\Omega_0$, and it defines the FEL parameter \cite{huang:PRSTAB2007}:
\begin{equation}\label{growth:1}
 \rho_0 = \left(\frac{a_0^2 n_0}{12 \pi \gamma_b^3 (1-\cos\theta)^2 }\right)^{1/3}\,,
\end{equation}
which corresponds to the power gain length, ${L_{g0} = \lambda_u/(4\pi\sqrt{3}\rho)}$.

The finite electron energy spread suppresses the amplification and modifies its spectral structure by narrowing the
resonance bandwidth and down-shifting the central wavelength as: 
\begin{equation}\label{resonance:0}
\omega_s\approx 2\gamma_b^2(\tilde{\omega}_e -\Omega_0)/(1+2\delta\gamma/\gamma_b)\,. 
\end{equation}
The maximal FEL parameter in this case decreases with the energy spread, and when $\delta\gamma/\gamma_b\gtrsim
\rho_0$, it can be estimated as $\rho\approx \sqrt{2/3} (\delta\gamma/\gamma_b)^{-1/2}\rho_0^{3/2}$.

The dependence of FEL parameter on the energy spread and signal frequency is presented in \cref{fig3}, where the
resonance condition \cref{resonance:0} is shown with the dashed white line. The maximal FEL parameter corresponding
to the chosen interaction parameters is $\rho_0 = 4.38\cdot 10^{-3}$.

\subsection{Account for the ellipticity of electron orbits}\label{sec2:2}

\begin{figure*}[ht!]\centering
\begin{tabular}{cc}\bf
 \bf (a) & \bf (b)\\
 \includegraphics[width=0.48\textwidth]{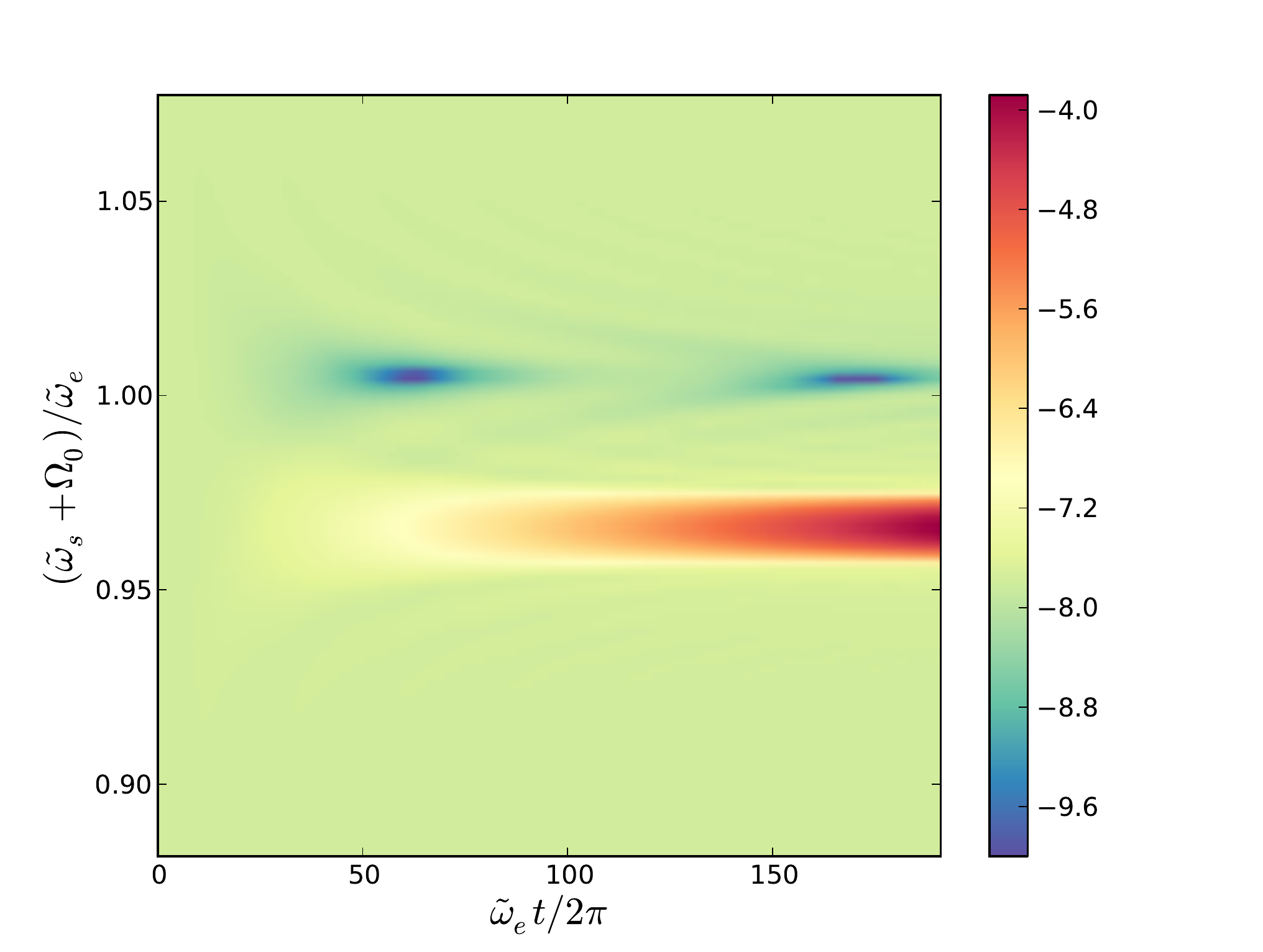}&
\includegraphics[width=0.48\textwidth]{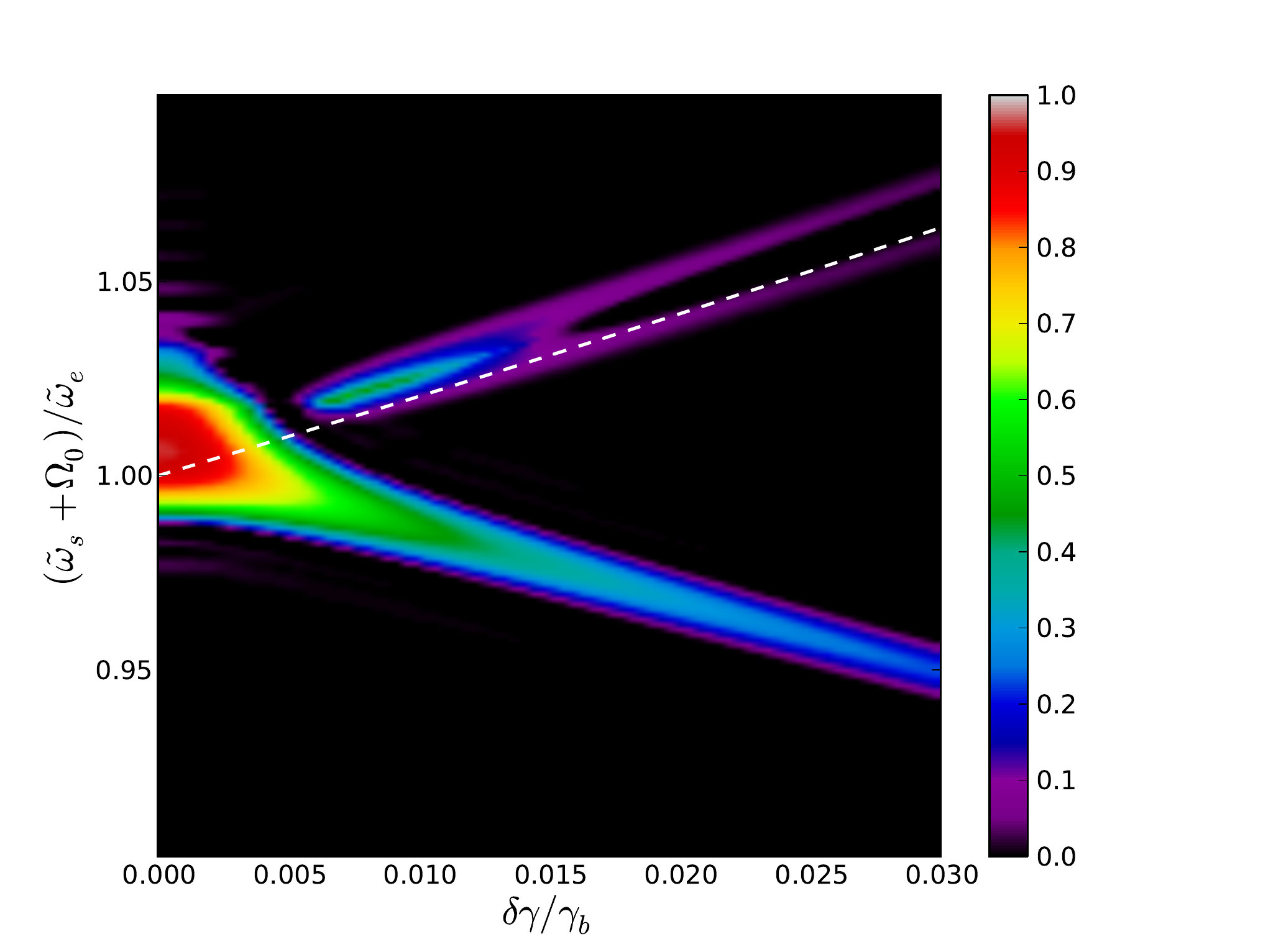}
\end{tabular}
\caption{ (a) Temporal dynamics of the signal wave as a function of its frequency calculated numerically. Color-scale is
logarithmic. (b) Normalized FEL parameter $\rho/\rho_0$ as a function of the longitudinal temperature and the signal
frequency calculated numerically for the elliptic electron mode.}\label{fig4}
\end{figure*}

For a more accurate analysis, one has to account for the dependence of electron oscillations frequency on the excursion
parameter $\xi$, which may be approximately described by \cref{ellipt_mode}. In this case, the analytical linearization
of
\cref{density:1,maxwell:0} would require further approximations and affect the accuracy of such description.
Alternatively, these equations may be solved numerically on a finite time interval via iterations. At each
iteration we integrate \cref{density:1,maxwell:0} over the time interval $(0,\tau_0)$, and written for the normalized
functions $N^{(n)} = a_0 n^*/(2\gamma^3\tilde{\omega}_s\tilde{\omega}_e)$ and $A^{(n)} = a_s/a_0$ in the following
form: 
\begin{subequations}\label{set3}
\begin{equation}\label{set3:1}
 N^{(n)} = -12\pi \rho_0^3 w \sin\chi \int_0^\tau \mathrm{d}\tau' A^{(n)}(\tau') \tilde{G}(\tau-\tau')\,,
\end{equation}
\begin{equation}\label{set3:2}
A^{(n+1)} = \frac{i}{\pi}\int_0^\tau \mathrm{d}\tau' \int_{-\pi/2}^{\pi/2} \mathrm{d}\chi N^{(n)}\,,
\end{equation}
\end{subequations}
where $\tau = \tilde{\omega}_e t$, $w = \tilde{\omega}_s/\tilde{\omega}_e$,
$O=\Omega_0/\tilde{\omega}_e$, 
$$\tilde{G} = \frac{\sin w\delta g (\tau)}{w\delta g} \mathrm{e}^{i (1-w)\tau} B(O,\tau,\chi)\,,$$ 
$\delta g = 2\delta\gamma/\gamma$, and the function $B$ is defined by \cref{ellipt_mode}. Considering a small
arbitrary ``seed'' $A^{(0)} = \mathrm{const}$, we calculate the corresponding $N^{(0)}$ from \cref{set3:1}, and then
integrate \cref{set3:2} to obtain $A^{(1)}$. 

The solution to \cref{set3} can be obtained by consequently repeating this procedure till the system converges,
$||A^{(n)}-A^{(n-1)}||\ll ||A^{(n)}||$. In practice, this algorithm converges rapidly, and can be efficiently used for
the parametric studies with complicated electron modes. This model also allows to include more effects, e.g. the
divergence of the electron beam along the non-trapped direction. To do so, in the integral in \cref{set3:1} we add 
the factor 
\begin{equation}\label{divergence}
n_0(t)/n_0(0) = \left(1+ (\delta v_y t/\sigma_y)^2 \right)^{-1/2},
\end{equation}
which describes a progressive decrease of electron density caused by the divergence. This modification we use
later when comparing with three-dimensional numerical tests.

Let us test the numerical model for the case of the elliptic electron mode (\ref{ellipt_mode}). In \cref{fig4}a, we plot
the evolution of the signal wave amplitude depending on the frequency for the same parameters as in \cref{fig3}, and the
energy spread $\delta\gamma/\gamma_b=0.01$. As expected the amplified mode grows exponentially and is down-shifted due
to the electron thermal motion. One can also see excitation of the non-resonant mode up-shifted due to the electron
temperature.

In \cref{fig4}b we reproduce numerically the map of the FEL parameter as a function of the frequency and energy spread
for the same parameters as in \cref{fig3}. One may see a remarkable agreement in \cref{fig3,fig4}b in most of the
features, and the maximal FEL parameter $\rho= 4.21 \cdot 10^{-3}$ in this calculation agrees with $\rho_0$ given by
\cref{growth:1}. The feature observed along,
$$\tilde{\omega}_s\simeq (\tilde{\omega}_e-\Omega_0)/(1-2\delta\gamma/\gamma_b)\,,$$ 
corresponds to a non-resonant up-shifted electron mode, and it is shown with the white dashed curve in \cref{fig4}b. The
center of the resonant region is slightly shifted, which results from the fact, that the frequency of elliptic electron
oscillations $\Omega'$ averaged over the distribution is lower than $\Omega_0$.

\section{Simulations with the free electron laser code}\label{sec3}

The theoretical model developed in the previous section describes the amplification at the linear stage, when the main
electron distribution is not affected by the growth of electron density perturbations. Therefore, such a model cannot
describe the instability saturation, which occurs when a significant part of the electrons are trapped in the
longitudinal potentials and stop transferring the energy to the signal wave. Near the saturation, the resonance
properties of the electron beam mode (e.g. electrons energy spread) may also be significantly modified. 

To account for the non-linear effects a self-consistent description is required. For this we use the particle free
electron laser code PLARES \cite{andriyash:JCP2014}. In this code the relativistic electrons are presented by the
macro-particles, and their unaveraged three-dimensional motion is coupled with the electromagnetic field through the
spectral Maxwell solver. The code may operate in the Cartesian or axisymmetric geometries, and for the present study we
use the Cartesian solver, which allows to model the asymmetric radiation profiles.

\begin{figure*}[ht!]\centering
\begin{tabular}{cc}\bf
 \bf (a) & \bf (b)\\
 \includegraphics[width=0.4\textwidth]{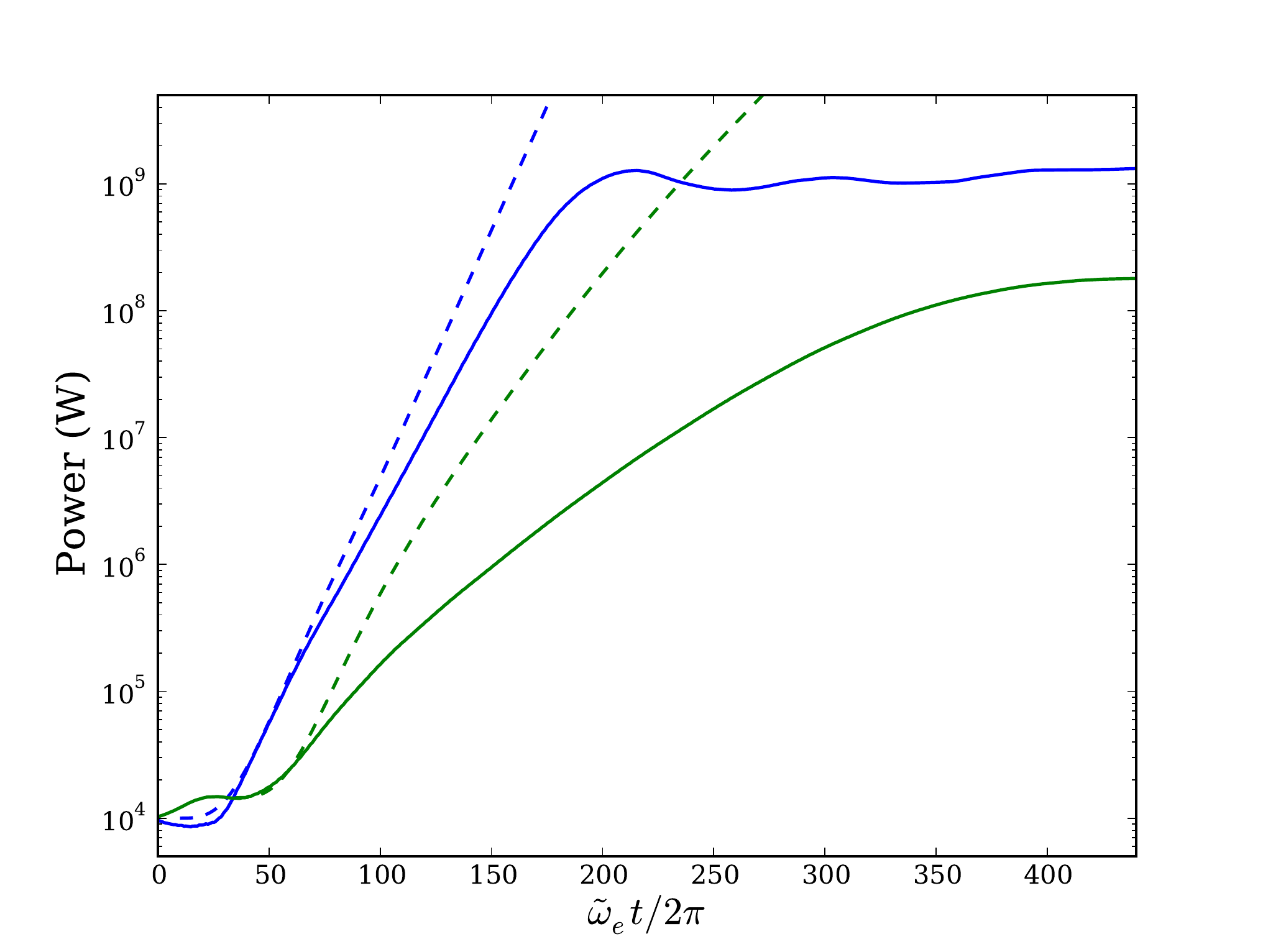}&
\includegraphics[width=0.6\textwidth]{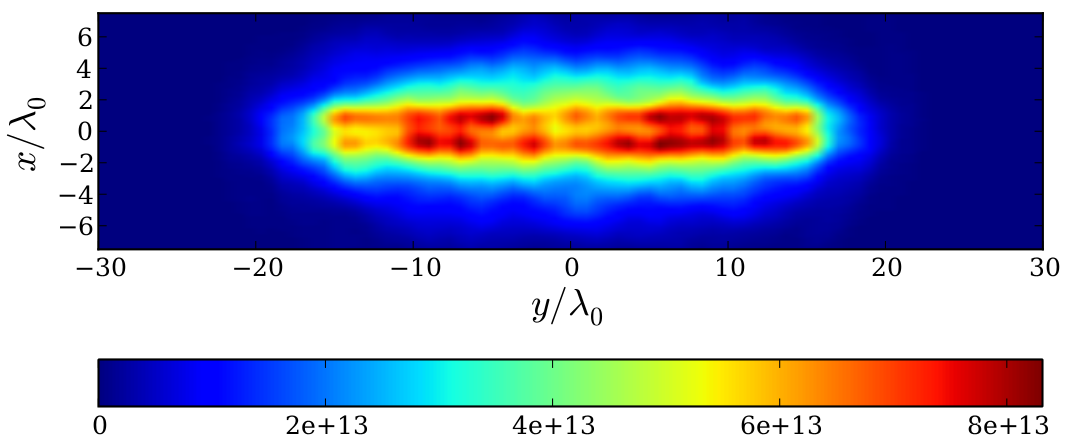}
\end{tabular}
\caption{Steady state simulations. (a) Radiation power as a function of time for the one-dimensional (blue) and
three-dimensional (green) cases modeled by PLARES (solid lines) and with help of \textit{ad hoc} iterative model
(\ref{set3}) (dashed lines). (b) The intensity profile of the X-rays at the end of the interaction in three-dimensional
simulation. The colorscale is in W/cm$^2$.}\label{fig5}
\end{figure*}


The general parameters in the numerical simulations are described in \cref{sec2:0}. In addition, to model the injection
of the electron beam into the lattice, we include a $300\lambda_0$ ($4.5\lambda_y$) linear ramp along
$z$-axis. The shot-noise in the system is completely suppressed and the simulations are seeded at the resonant
wavelength of 4.4 nm with a 1 kW pulse. In the three-dimensional case the transverse profiles of the seed are Gaussian
and correspond to a waist of 1.7 $\mu$m.

Let us first present the steady-state simulations, where the radiation field is considered as a single-frequency
infinitely long wave and it interacts with a small fraction of the electron beam in the periodic boundary conditions.
This case is close to the theoretical description developed in \cref{sec2}, however, it accounts for the dynamics of the
electron distribution and includes all related nonlinear effects. The results for the case of the monoenergetic electron
beam in one-dimensional (blue solid curve) and three-dimensional (blue solid curve) simulations are presented in
\cref{fig5}a. The results of PLARES simulations are compared with the description provided by the \textit{ad hoc} model
(\ref{set3}) (dashed curves), where the three-dimensional model also accounts for the divergence \cref{divergence}. For
the one-dimensional case, we see a good agreement for the early stage of amplification, while during the main stage the
growth rate is reduced by approximately 15 \% due to the modification of the electron beam main state. In the
three-dimensional simulation the amplification decreases even more due to the signal wave diffraction. The saturation
occurs at the distances $L_\mathrm{sat}=1$~cm and 2~cm and reaches relatively high powers at the 1 GW and 200 MW levels
for the 1D and 3D simulations, respectively.

The transverse profile of the radiation at the end of the 3D simulation is shown in \cref{fig5}b. One may clearly
observe that the resulting signal is stretched along the non-trapped direction due to the divergence of the electron
beam, and is also affected by the diffraction. The intensity modulations along the $x$-axis correspond to the
$\chi$ dependence of coupling between modes in \cref{dispersion:1}.

\begin{figure}[ht!]\centering
\includegraphics[width=0.48\textwidth]{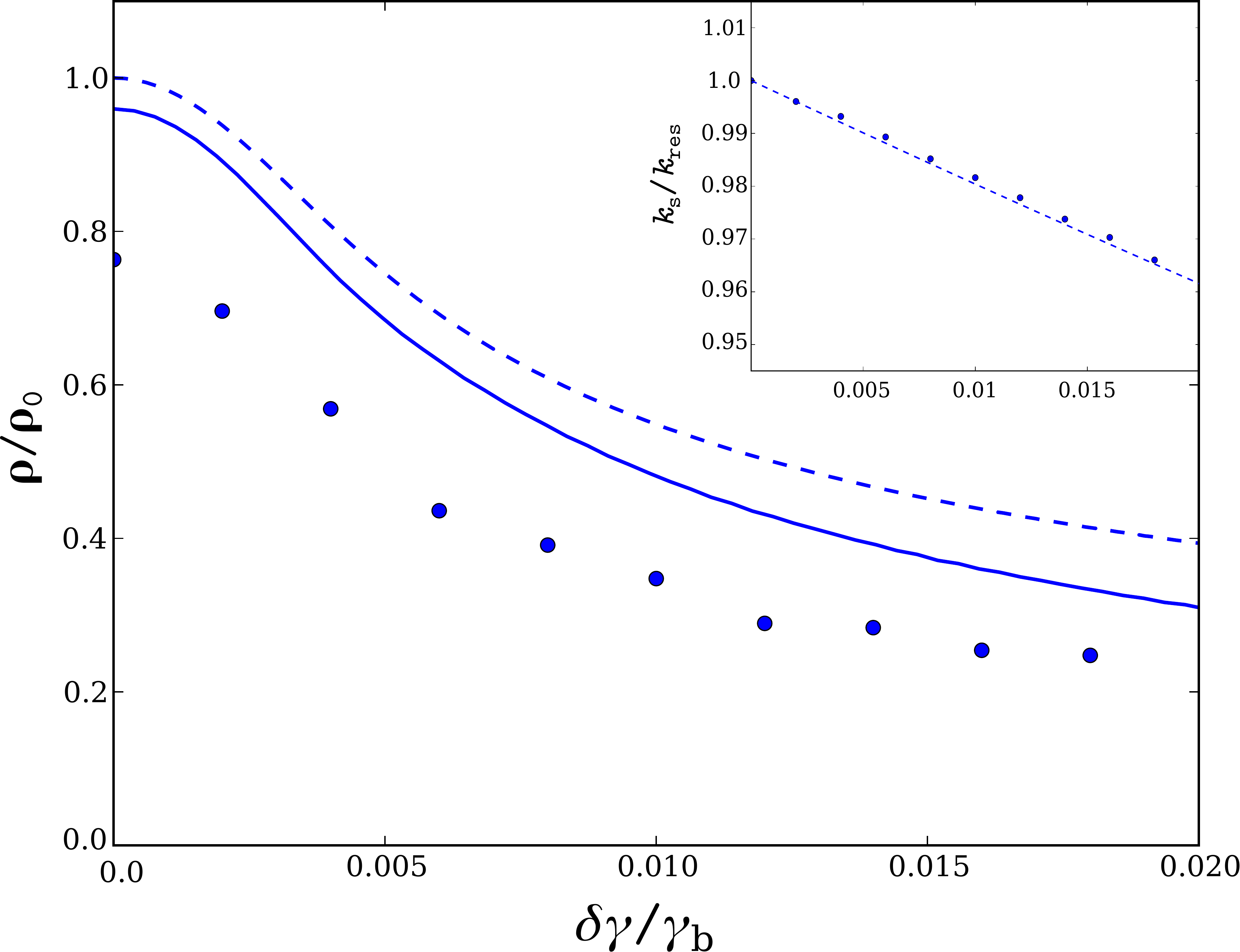}
\caption{Dependence of the FEL parameter on the energy spread of the electrons for the analytical (dashed line) and
numerical \textit{ad hoc} (solid line) models (\ref{set3}), and in the PLARES simulations (dots). The inset shows the 
dependence of the normalized resonant wavenumber on the energy spread in the simulations (dots) and according to
condition (\ref{resonance:0}) (dashed line).}\label{fig6}
\end{figure}

To study how the amplification is affected by electrons energy spread, we run a series of one-dimensional time-dependent
simulations. The term time-dependent here means that the spectral dynamics of the signal wave is modeled, and the
longitudinal profiles of the electron beam and of the seeding pulse are considered. In this model the resonance condition
appears naturally corresponding to the dominant mode. The 1.5 fs seed pulse with the Gaussian longitudinal profile is
used in the simulation, and its length is chosen to be much shorter than the length of the electron beam.

In \cref{fig6} we plot the FEL parameter as a function of $\delta\gamma/\gamma_b$ for the analytical model, numerical
solution model and PLARES simulations. The results indicate a good qualitative agreement in the instability behavior and
the demonstrate the quantitative difference of about 20 \% in the values, which is also observed in \cref{fig5}a. The
central wavenumbers corresponding to the growing mode in the simulations are shown in the inset of \cref{fig6} and
it confirms that the resonance follows the condition (\ref{resonance:0}). 

The results presented in this section prove a good quality of the theoretical model. The cases of more complex electron
distributions can be considered in the same way. The three-dimensional analysis indicates the importance of the
divergence of electron beam and the diffraction of the amplified radiation.


\section{Summary}\label{sec4}

We have presented a kinetic theory of the linear regime of the X-ray/XUV amplification in an electron beam, which
travels in a laser-induced transverse optical lattice. The theory accounts for the energy spread and angular divergence
of electron beam. For a simple realistic model of electron distribution the analytical expressions describe the
amplification rate and its dependence on the longitudinal energy spread of electrons. This point has a crucial
importance for the practical realizations of such a SR source. The presented results are confirmed by  numerical
simulations with the spectral FEL code PLARES, where the electron distribution is modeled via  particle methods and the
dynamics of electromagnetic field is modeled in a consistent way.


The presented analysis indicates that the amplification of the XUV light in the optical lattice is possible for the
electron parameters typical of  the state-of-the-art laser plasma accelerators. The account for the three-dimensional
effects evidences that the electron guiding in the optical lattice plays an important role by preserving the electron
flux. This opens the path for  practical realizations of the optical lattice XFEL without additional components for
electron focusing.

\section{Acknowledgements}

This work was partially supported by L'Agence Nationale de la Recherche (ANR) of France through the project LUCEL-X, and
by the European Research Council through the X-Five ERC project (Contract No. 339128).


\begin{thebibliography}{10}

\bibitem{emma:Nat2010}
P.~Emma et al.,
\newblock Nat. Phys. \textbf{4}, 641 (2010).

\bibitem{mcneil:NatPhot2010}
Brian W.~J. McNeil and Neil~R. Thompson,
\newblock Nat. Photon. \textbf{4}, 814 (2010).

\bibitem{albert:PRSTab2010}
F.~Albert et al.,
\newblock {Phys. Rev. ST Accel. Beams} \textbf{13}, 070704 (2010).

\bibitem{taphuoc:NatPhot12}
K.~Ta~Phuoc, S.~Corde, C.~Thaury, V.~Malka, A.~Tafzi, R.~C. Shah, S.~Sebban,
  and A.~Rousse,
\newblock Nat. Photon. \textbf{6}(5), 308 (2012)

\bibitem{huang:PRSTAB2007}
Zhirong Huang and Kwang-Je Kim,
\newblock Phys. Rev. ST Accel. Beams \textbf{10}, 034801 (2007).

\bibitem{sprangle:JAP1979}
P.~Sprangle and A.~T. Drobot,
\newblock J. Appl. Phys. \textbf{50}(4), 2652 (1979).

\bibitem{Bacci:PRSTAB2006}
A.~Bacci, M.~Ferrario, C.~Maroli, V.~Petrillo, and L.~Serafini,
\newblock Phys. Rev. ST Accel. Beams \textbf{9}(6), 060704 (2006).

\bibitem{sprangle:PRSTAB2009}
P.~Sprangle, B.~Hafizi, and J.~R. Pe\~nano,
\newblock Phys. Rev. ST Accel. Beams \textbf{12}(5), 050702 (2009).

\bibitem{esarey:RMP2009}
E.~Esarey, C.~B. Schroeder, and W.~P. Leemans,
\newblock Rev. Mod. Phys. \textbf{81}, 1229 (2009).

\bibitem{malka:POP2012}
V.~Malka,
\newblock Phys. Plasmas \textbf{19}(5), 055501 (2012).

\bibitem{faure:Nat2006}
J.~Faure, C.~Rechatin, A.~Norlin, A.~Lifschitz, Y.~Glinec, and V.~Malka.
\newblock Nature, \textbf{444}, 737--739 (2006).

\bibitem{kim:PRL2013}
Hyung~Taek Kim, Ki~Hong Pae, Hyuk~Jin Cha, I~Jong Kim, Tae~Jun Yu, Jae~Hee
  Sung, Seong~Ku Lee, Tae~Moon Jeong, and Jongmin Lee,
\newblock Phys. Rev. Lett. \textbf{111}, 165002 (2013).

\bibitem{leemans:PRL2014}
W.~P. Leemans, A.~J. Gonsalves, H.-S. Mao, K.~Nakamura, C.~Benedetti, C.~B.
  Schroeder, Cs. T\'oth, J.~Daniels, D.~E. Mittelberger, S.~S. Bulanov, J.-L.
  Vay, C.~G.~R. Geddes, and E.~Esarey,
\newblock Phys. Rev. Lett. \textbf{113}, 245002 (2014).

\bibitem{schlenvoigt:Nat2008}
H.-P. Schlenvoigt et al.,
\newblock Nat. Phys. \textbf{4}, 130 (2008).

\bibitem{fuchs:Nat2009}
Matthias Fuchs et al.,
\newblock Nat Phys. \textbf{5}, 826 (2009).

\bibitem{nakajima:NatPhys2008}
Kazuhisa Nakajima,
\newblock Nat. Phys. \textbf{4}, 92 (2008).

\bibitem{kapitza:MPCPS1933}
P.~L. Kapitza and P.~A.~M. Dirac,
\newblock Math. Proc. Cambridge \textbf{29}(02), 297 (1933).

\bibitem{bucksbaum:PRL88}
P.~H. Bucksbaum, D.~W. Schumacher, and M.~Bashkansky,
\newblock Phys. Rev. Lett. \textbf{61}(10), 1182 (1988).

\bibitem{Freimund:Nat2001}
D.L. Freimund, K.~Aflatooni, and H.~Batelaan,
\newblock Nature \textbf{413}, 132 (2001).

\bibitem{fedorov:APL1988}
M.~V. Fedorov, K.~B. Oganesyan, and A.~M. Prokhorov,
\newblock Appl. Phys. Lett. \textbf{53}(5), 353 (1988).

\bibitem{sepke:PRE2005}
S. Sepke, Y.Y. Lau, J.P. Holloway, and D. Umstadter,  
\newblock Phys. Rev. E \textbf{72}(2), 026501 (2005).

\bibitem{balcou:EPJD2010}
Ph. Balcou,
\newblock Eur. Phys. J. D \textbf{59}, 525 (2010).

\bibitem{andriyash:PRL2012}
I.~A. Andriyash, E.~d'Humi\`eres, V.~T. Tikhonchuk, and Ph. Balcou,
\newblock Phys. Rev. Lett. \textbf{109}, 244802 (2012).

\bibitem{andriyash:EPJD2011}
I.A. Andriyash, Ph. Balcou, and V.T. Tikhonchuk,
\newblock Eur. Phys. J. D \textbf{65}, 533 (2011).

\bibitem{frolov:NIMB2013}
E.N. Frolov, A.V. Dik, and S.B. Dabagov,
\newblock Nucl. Instrum. Meth. B \textbf{309}(0), 157 (2013).

\bibitem{andriyash:PRSTAB13}
I.~A. Andriyash, E.~d'Humi\`eres, V.~T. Tikhonchuk, and Ph. Balcou,
\newblock Phys. Rev. ST Accel. Beams \textbf{16}, 100703 (2013).

\bibitem{debus:APB2010}
A.D. Debus, M.~Bussmann, M.~Siebold, A.~Jochmann, U.~Schramm, T.E. Cowan, and
  R.~Sauerbrey,
\newblock Appl. Phys. B \textbf{100}(1), 61 (2010).

\bibitem{steiniger:NIMA2014}
K.~Steiniger, R.~Widera, R.~Pausch, A.~Debus, M.~Bussmann, and U.~Schramm.
\newblock Nucl. Instrum. Meth. A \textbf{740}(0), 147 (2014).

\bibitem{andriyash:JCP2014}
I.A. Andriyash, R.~Lehe, and V.~Malka,
\newblock J. Comput. Phys. \textbf{282}(0), 397 (2015).

\bibitem{gibbon:2005}
Paul Gibbon,
\newblock {\em Short Pulse Laser Interactions with Matter: An Introduction}.
\newblock Imperial College Press, 2005.

\bibitem{Kruer:1988}
William Kruer,
\newblock {\em The Physics of Laser Plasma Interactions}.
\newblock Westview Press, 1988.

\end{thebibliography}

\end{document}